\documentclass[12pt,epsfig,citesort]{article}
\usepackage{geometry}
\newcommand{\ice}[1]{\relax}
\usepackage{graphicx}
\usepackage{epsfig}
\usepackage{latexsym}
\usepackage{amsmath}
\usepackage{amsfonts}
\usepackage{amsxtra}

\def\be{\begin{equation}}
\def\ee{\end{equation}}
\def\bea{\begin{eqnarray}}
\def\eea{\end{eqnarray}}

\def\ap#1#2#3   {{ Ann. Phys. (NY)} {\bf#1} (#2) #3.}
\def\apj#1#2#3  {{  Astrophys. J.} {\bf#1} (#2) #3.}
\def\apjl#1#2#3 {{ Astrophys. J. Lett.} {\bf#1} (#2) #3.}
\def\app#1#2#3  {{ Acta. Phys. Pol.} {\bf#1} (#2) #3.}
\def\ar#1#2#3   {{ Ann. Rev. Nucl. Part. Sci.} {\bf#1} (#2) #3.}
\def\cpc#1#2#3  {{ Computer Phys. Comm.} {\bf#1} (#2) #3.}
\def\err#1#2#3  {{ Erratum} {\bf#1} (#2) #3.}
\def\ib#1#2#3   {{ ibid.} {\bf#1} (#2) #3.}
\def\jmp#1#2#3  {{ J. Math. Phys.} {\bf#1} (#2) #3.}
\def\ijmp#1#2#3 {{ Int. J. Mod. Phys.} {\bf#1} (#2) #3.}
\def\jetp#1#2#3 {{ JETP Lett.} {\bf#1} (#2) #3.}
\def\jpg#1#2#3  {{ J. Phys. G.} {\bf#1} (#2) #3.}
\def\mpl#1#2#3  {{ Mod. Phys. Lett.} {\bf#1} (#2) #3.}
\def\nat#1#2#3  {{ Nature (London)} {\bf#1} (#2) #3.}
\def\nc#1#2#3   {{ Nuovo Cim.} {\bf#1} (#2) #3.}
\def\nim#1#2#3  {{ Nucl. Instr. Meth.} {\bf#1} (#2) #3.}
\def\np#1#2#3   {{ Nucl. Phys.} {\bf#1} (#2) #3.}
\def\pcps#1#2#3 {{ Proc. Cam. Phil. Soc.} {\bf#1} (#2) #3.}
\def\pl#1#2#3   {{ Phys. Lett.} {\bf#1} (#2) #3.}
\def\prep#1#2#3 {{ Phys. Rep.} {\bf#1} (#2) #3.}
\def\prev#1#2#3 {{ Phys. Rev.} {\bf#1} (#2) #3.}
\def\prl#1#2#3  {{ Phys. Rev. Lett.} {\bf#1} (#2) #3.}
\def\prs#1#2#3  {{ Proc. Roy. Soc.} {\bf#1} (#2) #3.}
\def\ptp#1#2#3  {{ Prog. Th. Phys.} {\bf#1} (#2) #3.}
\def\ps#1#2#3   {{ Physica Scripta} {\bf#1} (#2) #3.}
\def\rmp#1#2#3  {{ Rev. Mod. Phys.} {\bf#1} (#2) #3.}
\def\rpp#1#2#3  {{ Rep. Prog. Phys.} {\bf#1} (#2) #3.}
\def\sjnp#1#2#3 {{ Sov. J. Nucl. Phys.} {\bf#1} (#2) #3.}
\def\spj#1#2#3  {{ Sov. Phys. JEPT} {\bf#1} (#2) #3.}
\def\spu#1#2#3  {{ Sov. Phys.-Usp.} {\bf#1} (#2) #3.}
\def\zp#1#2#3   {{ Zeit. Phys.} {\bf#1} (#2) #3.}

\def\beq{\begin{equation}}
\def\eeq{\end{equation}}
\def\bea{\begin{eqnarray}}
\def\eea{\end{eqnarray}}

\begin{document} 
\date{} 


\title{\bf Dark Matter and Yukawa Unification 
with Massive Neutrinos}
\author{ \bf  
M.E. Gomez $^{a}$,
S. Lola $^b$, 
P. Naranjo $^b$ 
and J. Rodriguez-Quintero $^{a}$
}

\maketitle

\begin{center}
$^{a)}$ Departamento de F\'{\i}sica Aplicada, University of Huelva, 
21071 Huelva, Spain \\
$^{b)}$ Department of Physics, University of Patras, 26500 Patras, Greece
\end{center}

\vspace*{2 cm}

\begin{quote}

\begin{abstract}
We revisit the WMAP dark matter constraints on Yukawa Unification 
in the presence of massive neutrinos. The large neutrino mixing indicated 
by the data modifies the predictions for the bottom quark mass,
and enables Yukawa  also for large $\tan\beta$, and for positive
$\mu$ that were  previously disfavoured.
As a result, the allowed parameter space
for neutralino  dark matter also increases, particularly 
for areas with resonant 
enhancement of the neutralino relic density.

\end{abstract}

\vspace*{5 cm}

{\it Proceedings of 4th International Workshop on the Dark Side of the Universe
(DSU 2008), Cairo,  June 2008. Published by American Institute of Physics.}



\end{quote}
\pagebreak


\section{Introduction}\label{intro}

Reconciling the Cold Dark Matter (CDM) predictions of 
supersymmetric models  with
the stringent constraints from the Wilkinson Microwave Anisotropy Probe
(WMAP) has been one of the challenges  within the particle physics community 
in recent years. The amount of CDM  deduced from WMAP data  
\cite{Bennett:2003bz,Spergel:2003cb} puts severe constraints on possible
Dark Matter Candidates, inluding the lightest supersymmetric particle (LSP). 
Additional constraints on the model parameters are obtained by
imposing Yukawa unification, and by taking into account the
bounds from Flavour Changing Neutral Currents (FCNC).

In addition to the above, the neutrino data of the past years
provided evidence for the existence of neutrino oscillations and masses,
pointing for the first time to physics beyond the Standard Model.
As expected, the additional interactions required to
generate neutrino masses also affect the energy dependence of the
couplings of the MSSM, and thus modify the Yukawa unification predictions.
A first observation had been that the additional interactions of
neutrinos, which affect the tau mass, may spoil bottom-tau Unification 
for small $\tan\beta$ \cite{VB}.
Subsequently, however, it has  been realised that large lepton mixing
naturally restores unification, and even allows Unification for intermediate
values of $\tan\beta$ that were previously disfavoured
\cite{LLR,CELW}. This is done by making the simple observation
that the $b-\tau$ equality at the GUT scale refers to the
$(3,3)$ entries of the charged lepton and
down quark mass matrices, while the detailed structure of the
mass matrices is not predicted by the Grand Unified
Group itself.
It is then possible to assume mass textures,
such that, after diagonalisation at   the
GUT scale, the $(m^{diag}_E)_{33}$ and $(m^{diag}_D)_{33}$
entries are no-longer equal.

In the current work, based on \cite{GLNR}, 
we revisit the issues of Dark Matter and Yukawa Unification
taking into account the effects of massive neutrinos and large
lepton mixing in See-Saw models, and 
extending previous results to large $\tan\beta$.
We find that the effects on the allowed parameter space are significant
and, in fact, it turns out that Yukawa Unification in the presence
of neutrinos is also compatible with a negative $\mu$, unlike what
happens if the effects of neutrinos are ignored. 
Passing to the relic density of neutralinos,
we study the consequences of large lepton mixing in 
the $\chi - \tilde{\tau}$ coannihilation region
and in resonances in the $\chi-\chi$ annihilation channels, finding sizeable 
effects, particularly in the latter case. It is interesting to note that for the cosmologically favoured area, it is also 
possible to observe tau flavour violation at the LHC, in the framework of non-minimal supersymmetric Grand Unification  \cite{edson}.

\section{Massive Neutrinos and Unification}

In the presence of massive neutrinos, the predictions for 
$m_b$ and unification clearly get modified. Radiative corrections  
from  the neutrino Yukawa couplings
have to be included in renormalisation group runs from 
$M_{GUT}$ to $M_{N}$ 
(scale of the heavy right-handed 
neutrinos).  Below $M_N$, right-handed neutrinos decouple from the 
spectrum and an effective  see-saw mechanism is operative; 
the relevant equations are given in \cite{GL}.
In addition, if the GUT scale 
lies significantly below a scale $M_X$, at which gravitational 
effects can no longer be neglected, the 
renormalization of couplings
at scales between $M_X$ and $M_{GUT}$ may induce 
additional effects to the running and the simplest example is
provided by minimal SU(5) \cite{Hisano} (however, 
modifications to soft masses are in this simplest case proportional 
to the $V_{CKM}$ mixing \cite{Hisano}, and thus are significantly suppressed).
Nevertheless, it has been realized  that the influence 
of the runs above the GUT scale on the Dark Matter abundance can be 
very sizeable \cite{Mamb}, due to  changes in
the relation between $m_{\tilde{\tau}}$ and $m_{\chi}$, 
which is crucial in the coannihilation 
area. This we discuss in a subsequent section.

In supersymmetric models, unification 
is  very sensitive to the model parameters \cite{pallis}, particularly the
Higgs mixing parameter, $\mu$. 
To correctly obtain  pole masses within this framework, 
the standard model and supersymmetric  threshold corrections
have to be included; for the bottom quark, these
corrections result to a $\Delta m_b$ that 
can be very large, particularly  for large values of $\tan\beta$ \cite{nath}.
Constraints from BR($b\rightarrow s \gamma$) 
are also included in the analysis.
Before passing to the results, however,
let us summarise a few facts on 
the possible range of the mass of the bottom quark:
the 2-$\sigma$ range for 
the  $\overline{MS}$ bottom running mass,
$m_b(m_b)$, is from  4.1-4.4 GeV. Moreover,
$\alpha_s(m_Z)=0.1172 \pm 0.002$, 
and the central  value of $\alpha_s$ corresponds 
to $m_b(m_Z)$ from  2.82 to 3.06 GeV.

\begin{figure}[!ht]
\hspace*{-0.3 cm}
\begin{center}
\includegraphics[width=8cm,height=7cm]{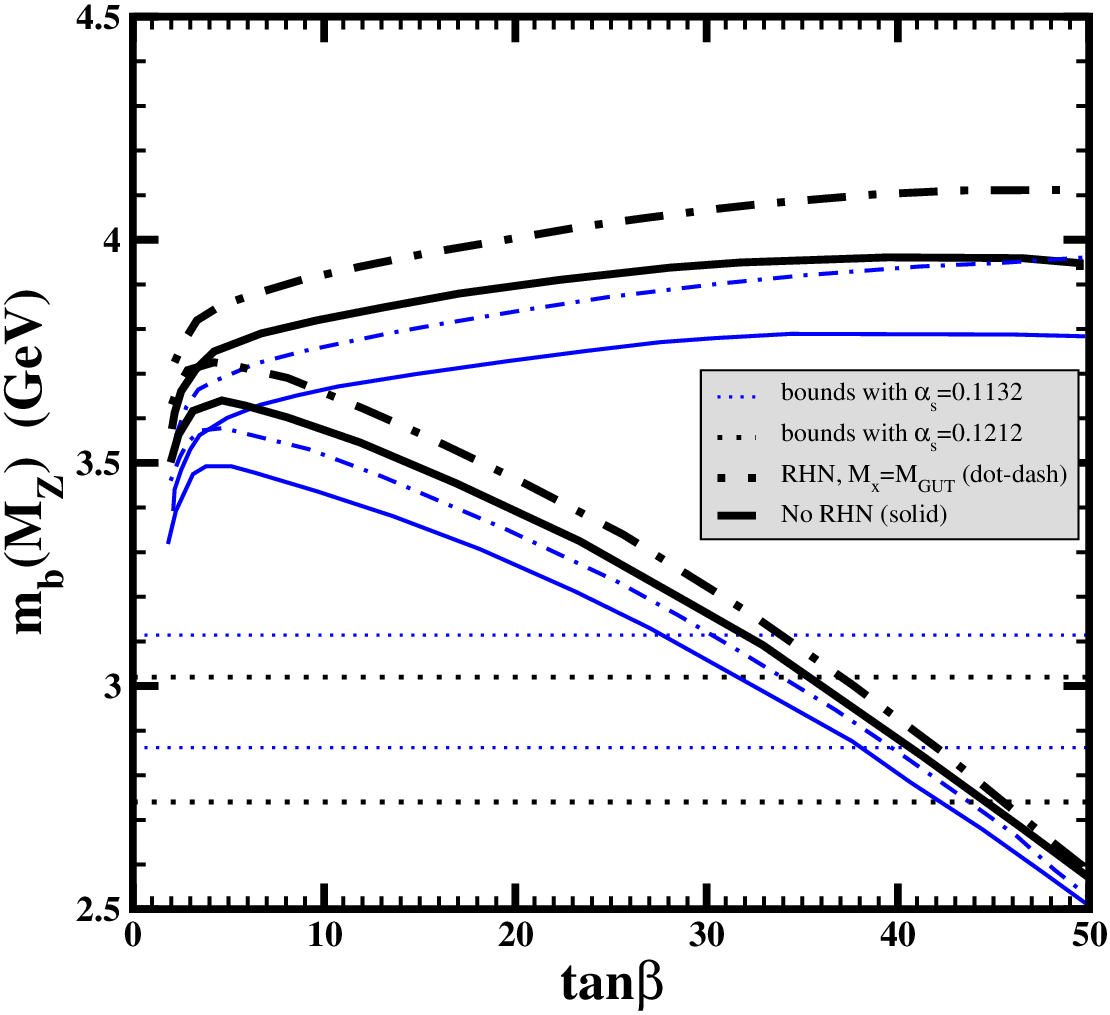}
\caption{ \it 
The value of $m_b(M_Z)$ versus $\tan \beta$ assuming
$\lambda_b = \lambda_\tau$ at the high scale in the absence 
of lepton mixing. We work with
the following set of parameters:
$m_{1/2}=800$ GeV , $A_0=0$ GeV, $m_{0}=600$ GeV. The experimental range of $m_b$ 
(horizontal lines) is also shown, the black thick
(blue thin) lines have $\alpha_s$ = 0.1212 (0.1132), the dot-dash lines  
are obtained within the MSSM , the solid lines include neutrino Dirac Yukawa up to the 
scale $M_N=3\times 10^{14}$. The upper (lower) set of
lines corresponds to $\mu>0\,\,(\mu<0)$.
}
\end{center}
\label{mb1}
\end{figure}

In Fig. \ref{mb1}, we summarize the predictions for $m_b$
in mSUGRA and in the presence of massive neutrinos. In order to discuss 
the dependence of $m_b(M_Z)$ on $\tan\beta$, 
we consider the set of soft parameters: $m_{\frac{1}{2}}=800$ GeV, $A_0=0$, $m_0=600$ GeV. The figure exhibits the well-known 
fact  that in the absence of phases or
large trilinear terms,  $\Delta m_b$ is positive for $\mu$ positive,
and therefore the theoretical prediction for the $b$ quark pole mass is 
too high to be reconciled with $b-\tau$ 
unification. On the other hand, 
for $\mu<0$,  $\Delta m_b$ is negative and the theoretical prediction 
for the b quark mass can lie within the experimental range 
for values of $\tan\beta$ between roughly 25 and 45;
clearly, for a large $\tan\beta$ it is mandatory
to take into account
the large supersymmetric corrections to $m_b$ 
~\cite{Hal,CW}.  

The analysis in the presence of massive neutrinos takes into account 
only the third generation couplings , from the $M_{GUT}$ to the
scale of the right handed neutrino masses, 
and evolve the light neutrino
mass operator from this scale down to $M_Z$. 
A large value  of  the Dirac-type 
neutrino Yukawa coupling, $\lambda_N$ at the GUT scale
may arise naturally within the framework of Grand Unification, and
its value is determined by demanding a third generation
low energy neutrino mass of $m_{\nu_3}=0.05$~eV.
The predictions for $m_b(M_Z)$ using the 
lower and upper bounds of the 2-$\sigma$ experimental range of 
$\alpha_s$ and the correponding range for $m_b(M_Z)$ after the evolution of the
bounds on $m_b(m_b)$ are shown for a scale $M_N = 3\times 10^{14}$~GeV.

We observe that for $\mu>0$ the prediction 
for $m_b(M_Z)$ is always very large, despite its dependence on the 
soft terms through $\Delta m_b$. For the values of the soft terms considered in  Fig.{\ref{mb1}}, the allowed 
range of $\tan\beta$ shrinks from $27-44$ to $30-45$ when we introduce 
the effect of see-saw neutrinos. It is also shown that the influence 
of  runs above $M_{GUT}$ is too small to have any significant effect.

The results are significantly modified once we consider the effects of lepton mixing
in the diagonalisation and running of couplings from high to low energies. In order to
show this, we focus on $b-\tau$ 
unification within the framework of SU(5) gauge unification and flavour symmetries that
provide consistent patterns for mass and mixing hierarchies, and naturally
reconcile a small $V_{CKM}$ mixing with a large charged 
lepton one. Taking into account the particle content of SU(5) representations
(with symmetric up-type mass matrices, and down-type mass matrices that are transpose to
the ones for charged leptons), 
one finds the approximate relations
\begin{equation}
{\mathcal {M}}_d^0 \propto A\,\left(\begin{array}{cc}
0 & 0 \\
x & 1  
\end{array}\right), \,\,\,\,
{\mathcal {M}}_{\ell}^0 \propto A\,\left(\begin{array}{cc}
0 & x \\
0 & 1 
\end{array}\right)
\label{Ml}
\end{equation}
which, after diagonalization, lead to 
\begin{equation}
\frac{m_b^0}{1+x^2}=\frac{m_{\tau}^0}{1-x^2}\,\rightarrow \,
m_b^0=m_{\tau}^0\,\left(1-\underbrace{2x^2}_{\delta}+{\mathcal{O}}
\left(\delta ^2\right)\right)
\label{btau}
\end{equation}
where $\delta$ parametrises  the flavour mixing in the (2,3) sector. 

\begin{figure}[!ht]
\hspace*{-0.3 cm}
\includegraphics[width=8cm,height=7cm]{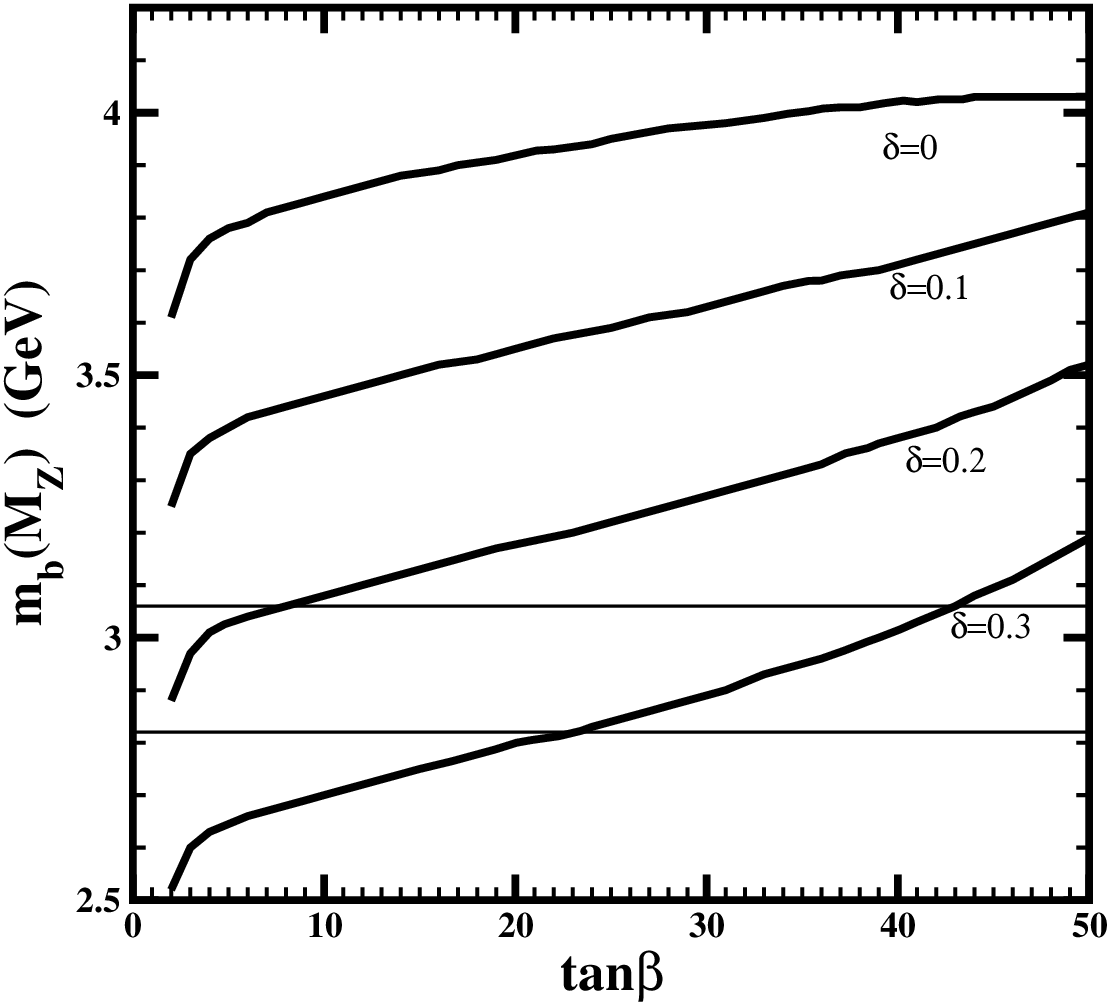}
\includegraphics[width=8cm,height=7cm]{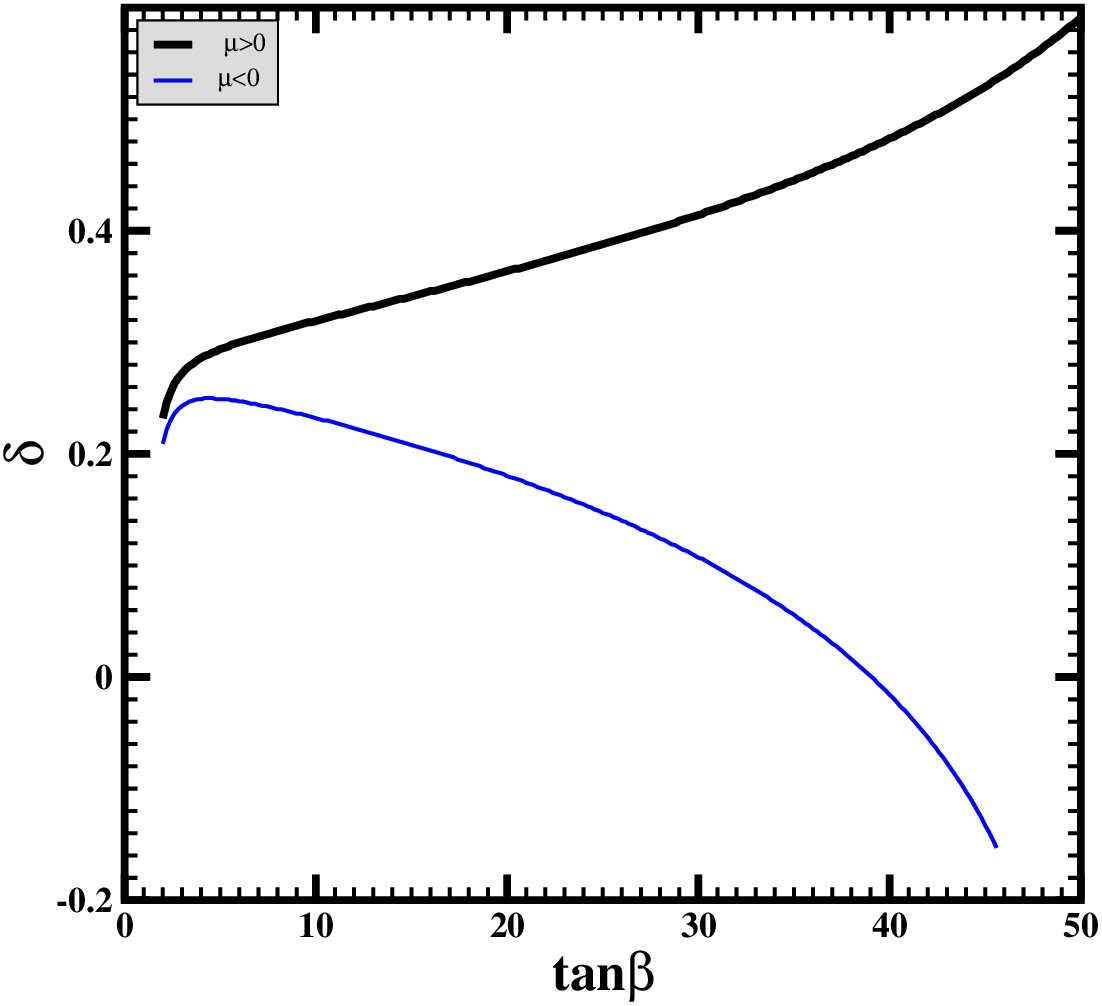}
\caption{\it 
In the left panel, we show $m_b$ as a function of $\tan\beta$, when including
lepton mixing effects. In the right panel, we show the required values
$\delta$ for consistent unification. We use the set of 
soft terms of Fig.\ref{mb1}, $\alpha_s=0.1172$ and impose $m_b(M_Z)=2.92$~GeV. The 
upper (lower) line corresponds to $\mu>0$ ($\mu<0$).
 }
\label{deltb}
\end{figure}

In the left  pannel of Fig.\ref{deltb} we show the change of $m_b$ as a 
function of $\tan\beta$, when the effects from large lepton mixing are appropriately considered. Comparing with the previous plots,  we 
see how solutions with positive
$\mu$ are now viable, for the whole range of $\tan\beta$.
The appropriate size of the parameter $\delta$ in each case 
can be determined by imposing 
the relation $\lambda_\tau=\lambda_b(1+\delta)$ at $M_{GUT} $ and investigating
the values that are required in order to obtain a correct prediction for  $m_b(M_Z)$.
This is shown in the right pannel of Fig.\ref{deltb}, where we demand 
a value of $m_b(M_Z)$  at
the center of its experimental range, for the central value of $\alpha_s$.

\section{Dark Matter constraints and Yukawa unification}

In mSUGRA (or the CMSSM) for choices of soft terms below the 
TeV scale, the LSP is Bino like and the prediction for $\Omega_\chi h^2$ is 
typically too large for models that satisfy the experimental constraints on 
SUSY. In fact, the values of WMAP can essentially  be obtained in two regions:
\begin{itemize}
\item $\chi-\tilde{\tau}$ coannihilation region that occurs for 
$m_\chi\sim m_{\tilde{\tau}}$.
\item Resonances in the $\chi-\chi$ annihilation channel, which occur 
for  $m_A\sim 2 m_\chi$.
\end{itemize} 

Since the above areas are ``fine-tuned'',  they will inevitably be 
sensitive to the changes induced by GUT unification and sizeable mixing in
the charged lepton sector. 
The runs corresponding to 
$M_X>M_{GUT}$ have a big impact on the 
neutralino relic density. The large
values of the gauge unified coupling $\alpha_{SU(5)}$ tend to 
increase the values
of   $m_{\tilde{\tau}}$, even if we start with small $m_0$ at $M_X$.

\begin{figure}[!ht]
\centering
\includegraphics[width=7.5cm,height=7cm]{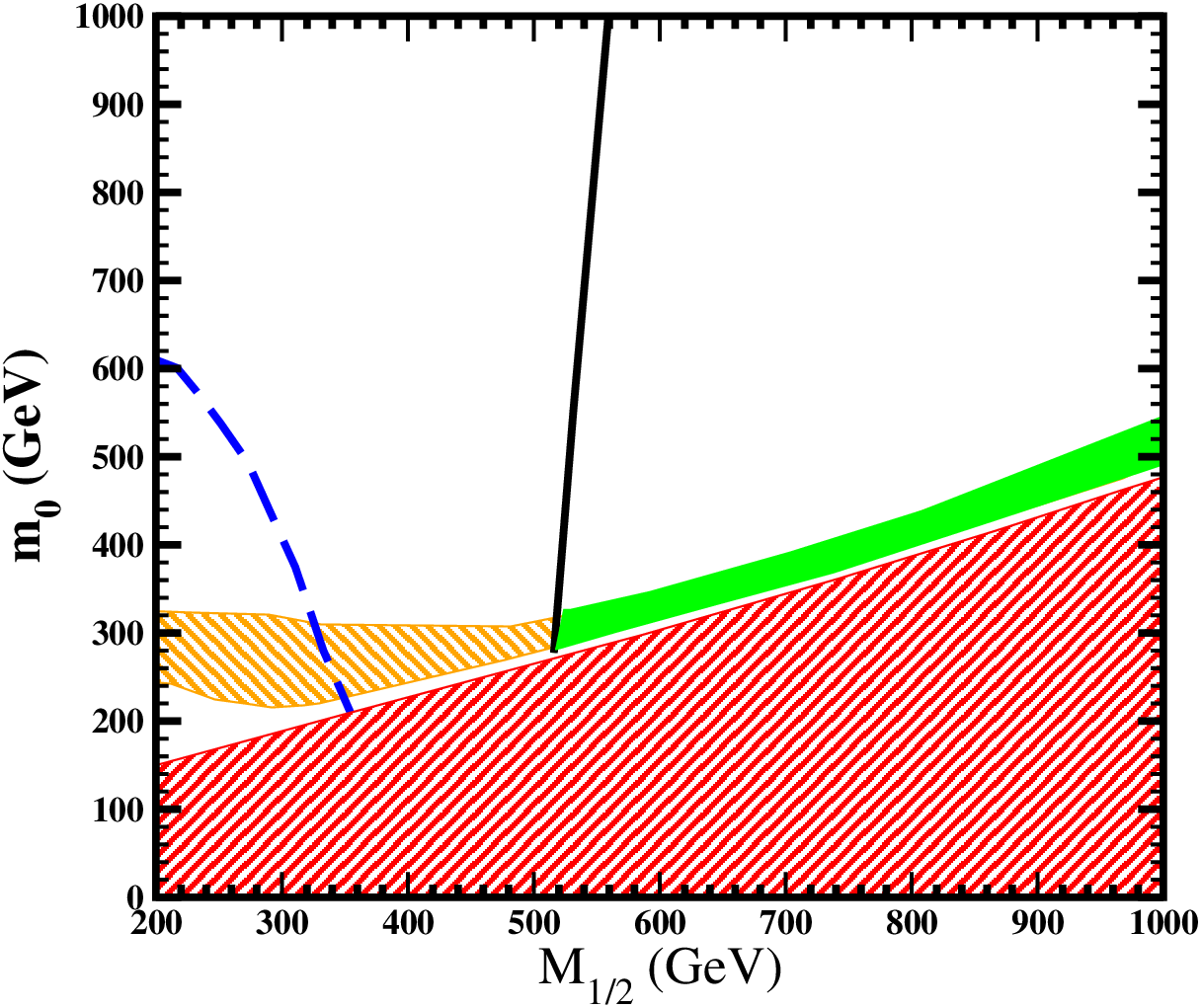}
\includegraphics[width=7.5cm,height=7cm]{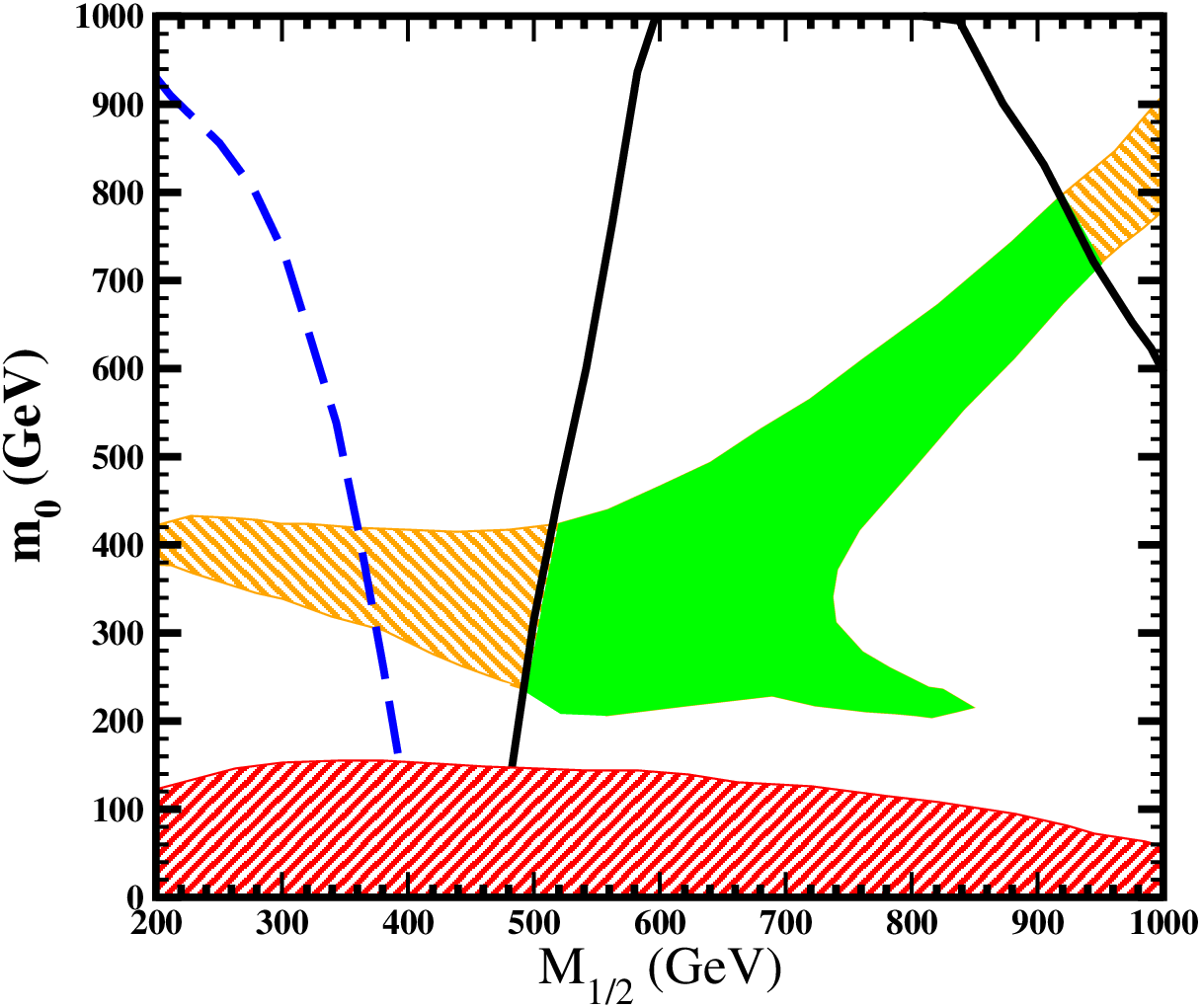}
\caption{\it WMAP allowed area (green) for the case 
of $\tan \beta=45$, $\mu>0$, $A_0=m_0$, $m_b\left(M_Z\right)=2.92$ GeV and 
$\delta \sim 0.42$ for the same set of parameters as in Fig.\ref{mb1}, 
without (left) and with (right) the $SU(5)$ running. The solid (dash) line corresponds to $m_h=114$~GeV (BR($b\rightarrow s \gamma$)=$2.8\cdot 10^{-4}$). In 
the lower (red) area the LSP is a stau.  }
\label{area45}
\end{figure}

\begin{figure}[!ht]
\centering
\includegraphics[width=7.5cm,height=7cm]{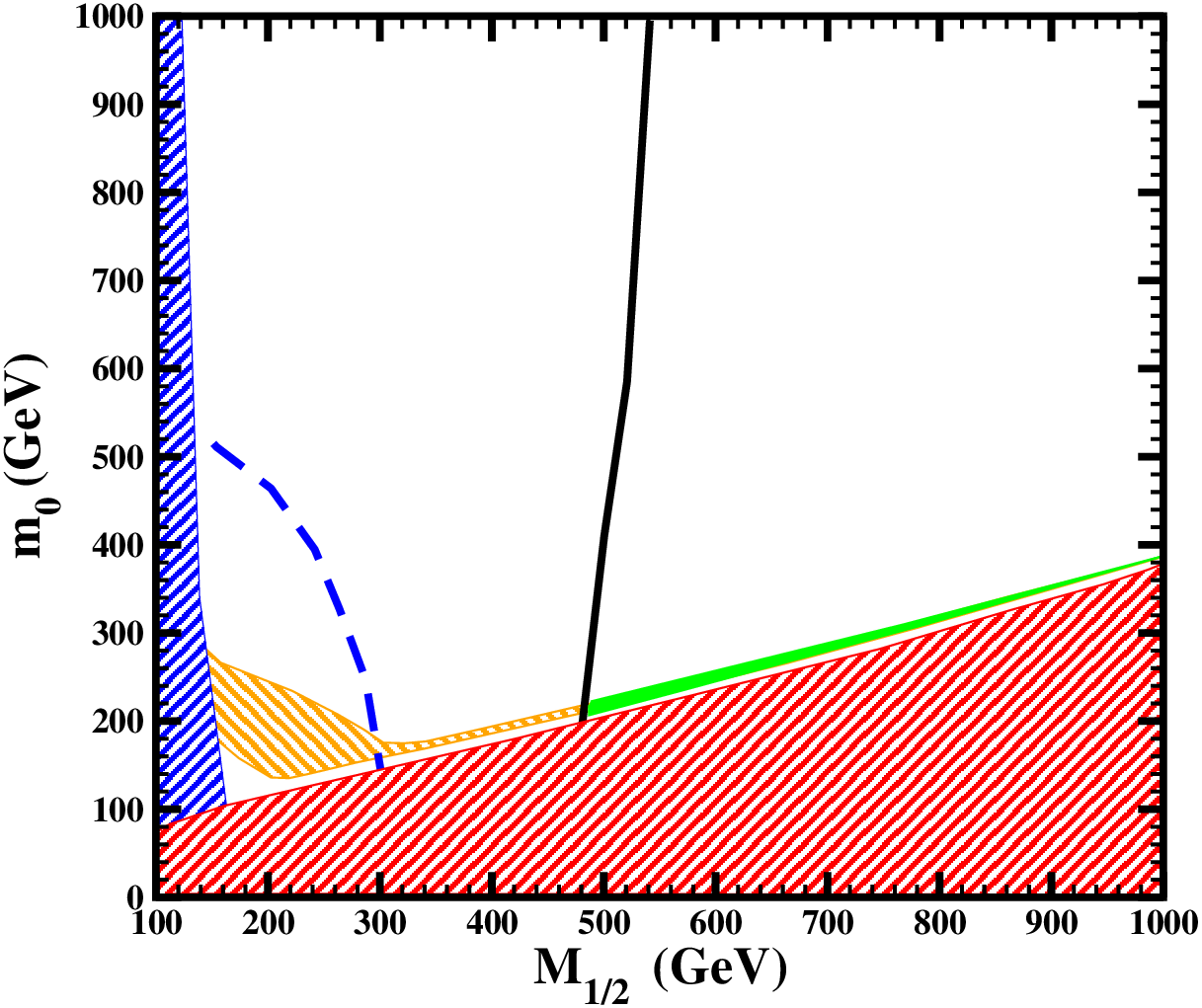}
\includegraphics[width=7.5cm,height=7cm]{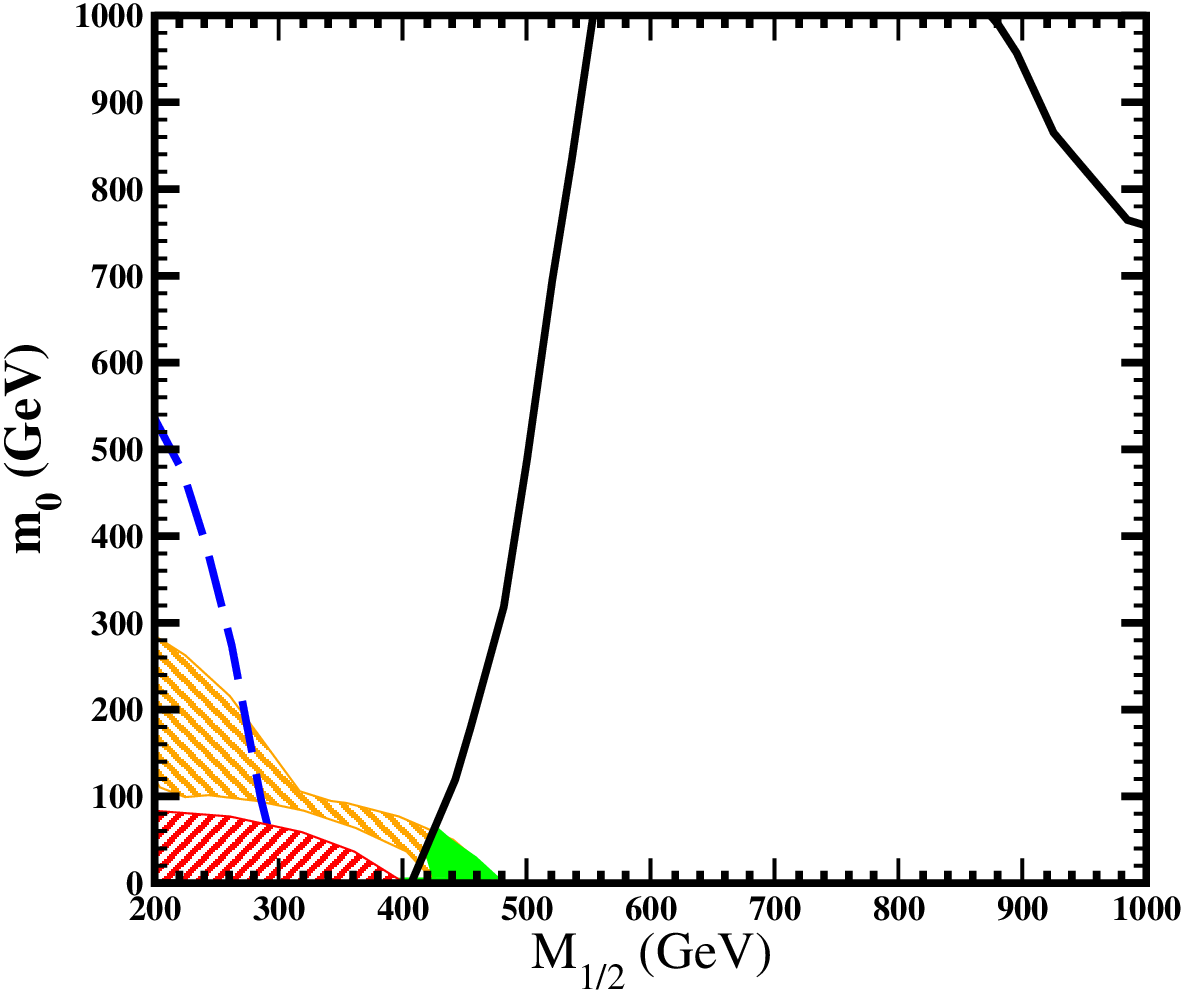}
\caption{\it The same pair of plots as in Fig.\ref{area45} but for 
$\tan \beta=35$~\cite{edson}. Here, the parameter $\delta\sim 0.37$.}
\label{area35}
\end{figure}

We see that the consideration of mixing effects, in combination with 
the inclusion of effects from the runs above $M_{GUT}$, significantly enhances 
the allowed parameter space (green area), an effect that is more visible
for large $\tan\beta$. Areas with different colours are excluded \cite{GLNR}. The reduction of the allowed parameter space for smaller values of $\tan\beta$, is already evident for $\tan\beta =35$. The  
$tan\beta=35$ picture has been discussed in \cite{edson}, where it was shown that
the WMAP allowed region can be compatible with observable  flavour violation at the LHC. 

The case with $\mu<0$ and a more detailed discussion of lepton mixing
effects are presented in \cite{GLNR}. 

\section{Conclusions}

We revisited the WMAP dark matter constraints on Yukawa Unification 
in the presence of massive neutrinos. Large neutrino mixing, as indicated 
by the data modifies the predictions for the bottom quark mass,
and enables Yukawa  also for large $\tan\beta$ and for positive
$\mu$ that were  previously disfavoured.
A direct outcome is that the allowed parameter space
for neutralino  dark matter also increases, particularly 
for areas with resonant enhancement of the 
neutralino relic density.

For completeness, we also note that for the cosmologically favoured parameter
region, we found lepton flavour violating rates very close to the current
experimental bounds \cite{new}. 
Finally, interesting effects may arise in the case of non-universal
soft terms. These are also discussed in detail in \cite{new}.

\vskip 1. cm
~\\
{\bf Acknowledgements} 
The research of S. Lola and P. Naranjo is funded by the FP6 
Marie Curie Excellence
Grant MEXT-CT-2004-014297.  The work of M.E.G, C.P and J.R.Q is supported by the Spanish MEC
projet FPA2006-13825 and the projet P07FQM02962 funded 
by ``Junta de Andalucia''.


\begin{thebibliography}{99}

\bibitem{Bennett:2003bz}
C.~L.~Bennett {\it et al.},
Astrophys.\ J.\ Suppl.\  {\bf 148}, 1 (2003)

\bibitem{Spergel:2003cb}
D.~N.~Spergel {\it et al.}  [WMAP Collaboration],
Astrophys.\ J.\ Suppl.\  {\bf 148}, 175 (2003)



\bibitem{VB}
F.~Vissani and A.~Y.~Smirnov,
  Phys.\ Lett.\  B {\bf 341}, 173 (1994),
A.~Brignole, H.~Murayama and R.~Rattazzi,
  Phys.\ Lett.\  B {\bf 335}, 345 (1994)

\bibitem{LLR}
  G.~K.~Leontaris, S.~Lola and G.~G.~Ross,
  Nucl.\ Phys.\  B {\bf 454} (1995) 25.

\bibitem{CELW}
M. Carena, J. Ellis, S. Lola and C. Wagner, 
Eur. Phys. J. C{\bf12} (2000) 507.


\bibitem{GLNR}
 M.~E.~Gomez, S.~Lola, P.~Naranjo and J.~Rodriguez-Quintero,
  arXiv:0901.4013 [hep-ph].

\bibitem{edson}
  E.~Carquin, J.~Ellis, M.~E.~Gomez, S.~Lola and J.~Rodriguez-Quintero,
  arXiv:0812.4243 [hep-ph].


\bibitem{GL}
B.~Grzadkowski and M.~Lindner,
Phys.\ Lett.\ B {\bf 193} (1987) 71;
Yu.~F.~Pirogov and O.~V.~Zenin,
Eur.\ Phys.\ J. C {\bf 10} (1999) 629;
N.~Haba, N.~Okamura and M.~Sugiura,
Prog.\ Theor.\ Phys.\ {\bf 103} (2000) 367,
Eur.\ Phys.\ J. C {\bf 10} (1999) 677.



\bibitem{Hisano}
J.~Hisano and D.~Nomura,
Phys.\ Rev.\ D {\bf 59}, 116005 (1999).

\bibitem{Mamb}
L.~Calibbi, Y.~Mambrini and S.~K.~Vempati,
  JHEP {\bf 0709}, 081 (2007).
\bibitem{pallis}
 M.~E.~Gomez, G.~Lazarides and C.~Pallis,
  Nucl.\ Phys.\  B {\bf 638}, 165 (2002), 
  Phys.\ Rev.\  D {\bf 67}, 097701 (2003).



\bibitem{nath}
  M.~E.~Gomez, T.~Ibrahim, P.~Nath and S.~Skadhauge,
  Phys.\ Rev.\  D {\bf 70}, 035014 (2004), 
  Phys.\ Rev.\  D {\bf 72}, 095008 (2005), 
  Phys.\ Rev.\  D {\bf 74}, 015015 (2006).



\bibitem{Hal}
 L.~J.~Hall, R.~Rattazzi and U.~Sarid,
  Phys.\ Rev.\  D {\bf 50}, 7048 (1994)

\bibitem{CW}
M.~S.~Carena, M.~Olechowski, S.~Pokorski and C.~E.~M.~Wagner,
  Nucl.\ Phys.\  B {\bf 426}, 269 (1994)


\bibitem{new}
 M.~E.~Gomez, S.~Lola, P.~Naranjo and J.~Rodriguez-Quintero, in preparation.



\end{thebibliography}
\end{document}